\documentclass[12pt]{article}

\usepackage[totalheight = 23cm, totalwidth = 17cm]{geometry}
\usepackage{amssymb,amsmath,amsfonts,amsbsy,epsfig,psfrag}

\begin{document}

\title{On two string axions as inflaton}

\author{
Jinn-Ouk Gong\footnote{jgong@kasi.re.kr}
\\ \\
\textit{International Center for Astrophysics}
\\
\textit{Korea Astronomy and Space Science Institute}
\\
\textit{Daejeon, Republic of Korea}
}

\date{\today}

\maketitle

\begin{abstract}

We study an inflationary model where two decoupled string axions drive inflation.
The number of $e$-folds is dependent on the energy scale and the decay constant, but
is almost independent of the angular component in spite of the rich geometry of the
moduli space. This suppresses the nearly scale-invariant spectrum of the
isocurvature perturbations, making the power spectrum of the primordial density
perturbations dominated by the adiabatic component. We also briefly discuss some
amendments to improve the situation.

\end{abstract}

\thispagestyle{empty}
\setcounter{page}{0}
\newpage
\setcounter{page}{1}

\section{Introduction}

Inflation \cite{inf} is currently the most promising candidate to solve many
cosmological problems. It is, however, still unclear yet how to implement a
consistent and successful scenario in the context of string theory. Fortunately,
after the de Sitter vacua construction in type IIB theory \cite{kklt}, there have
been considerable advances to this end for recent years \cite{advances}. An inflaton
candidate of particular interest is the string axion\footnote{See, e.g. Refs.
\cite{axionreview} for detailed accounts on the cosmological aspects of axion.},
which is known to exist abundantly in string theory, and they have flat potentials
even after all the moduli are stabilised.

In this paper, we study a simple inflation model which consists of two axion fields.
Since there are plenty of axions, it is not difficult to construct an inflation
model with the number of $e$-folds far larger than 60 using many fields. As
inflation proceeds, however, fewer and fewer fields are responsible for inflation
because heavy fields drop out of the inflationary regime and contribute no more.
Thus, it is very conceivable that in the final stage of inflation, i.e. the last 60
$e$-folds relevant for the observable universe, significantly smaller number of
fields drive inflation. In this regard, it is natural to consider the simplest
two-field model first which is considered important in building inflation models
with multiple fields \cite{2fields}.


This paper is outlined as follows. In Section~\ref{secpotential} we briefly review
the potential of two axion fields. In Section~\ref{secinflation} we study the
dependence of the number of $e$-folds on the parameters of the potential and the
density perturbations produced during inflation. In Section~\ref{secremedies} we
discuss how the problems associated with the simplest model could be alleviated. In
Section~\ref{summary} we give a brief summary.

\section{Potential}
\label{secpotential}

\begin{figure}[h]
\psfrag{fphi}{$\phi/f_\phi$}%
\psfrag{fpsi}{$\psi/f_\psi$}%
\begin{center}
\includegraphics[width = 0.5\linewidth]{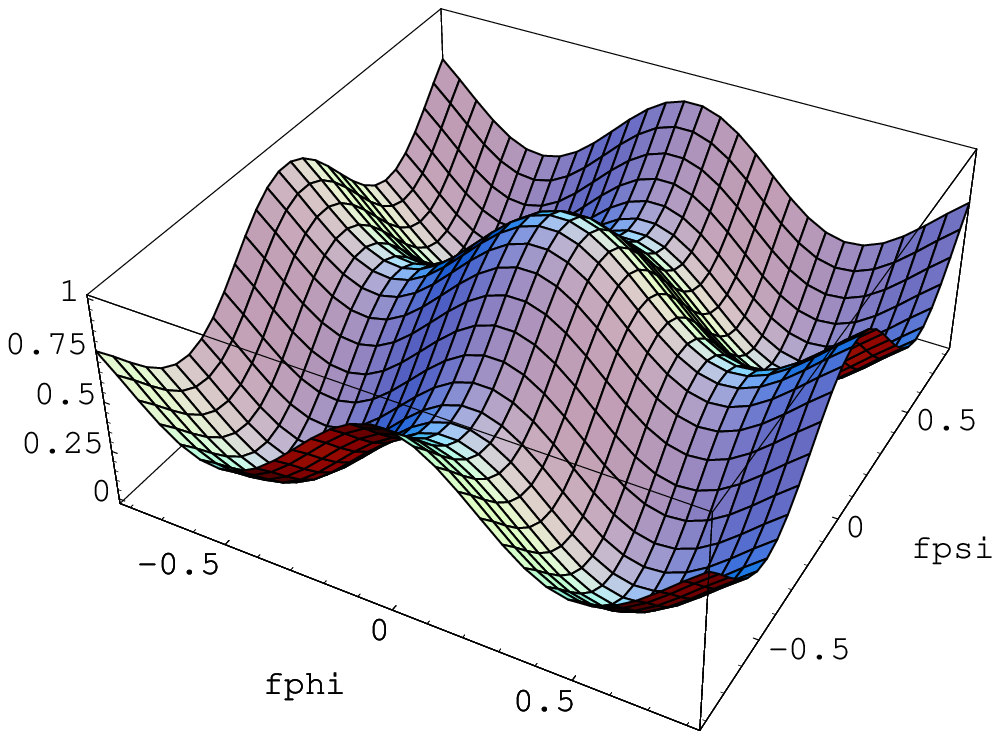}
\includegraphics[width = 0.4\linewidth]{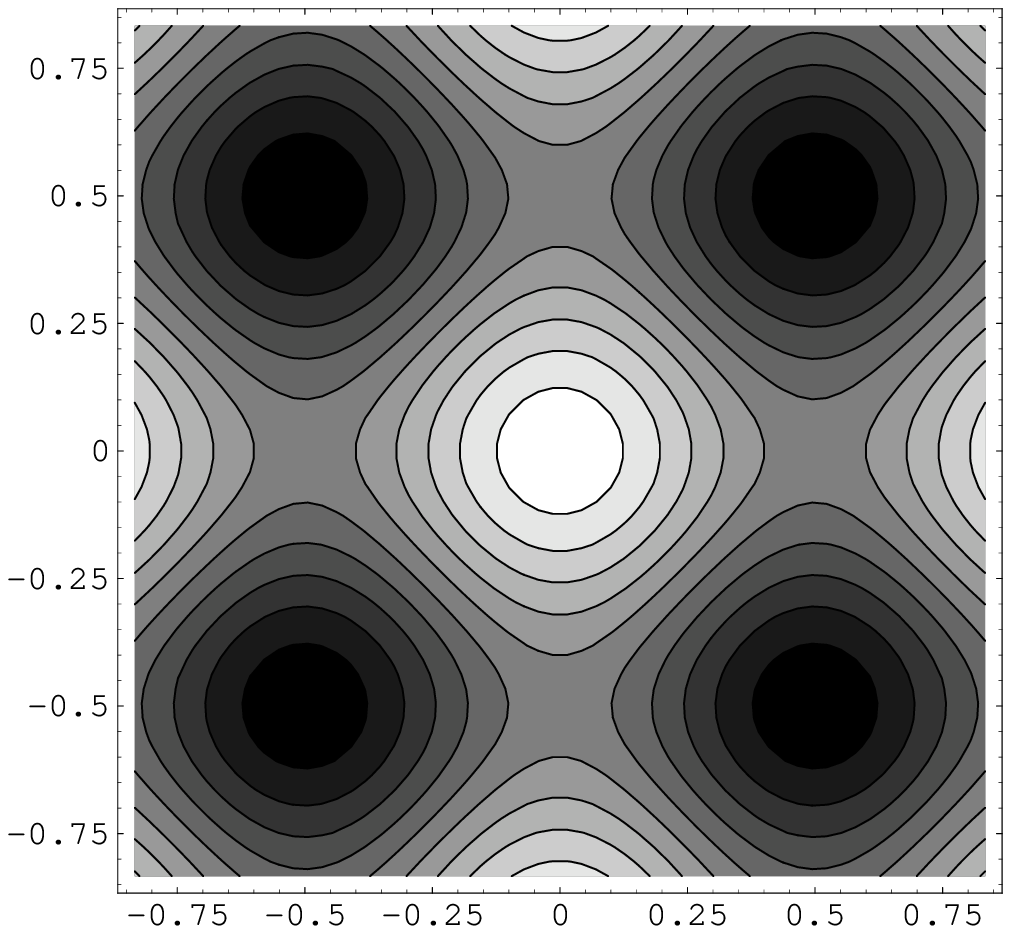}
\end{center}
\caption{(Left) plot of the potential, rescaled to $V_\mathrm{max} = 1$,
and (right) contour plot of the same region.}%
\label{figpotential}
\end{figure}

As mentioned in the previous section, we will be interested in the two-field case,
so the potential is written as
\begin{equation}\label{potential}
V(\phi,\psi) = \Lambda_\phi^4 \left[ 1 + \cos \left( \frac{2\pi\phi}{f_\phi} \right)
\right] + \Lambda_\psi^4 \left[ 1 + \cos \left( \frac{2\pi\psi}{f_\psi} \right)
\right] \, ,
\end{equation}
where $\Lambda_\phi$ and $\Lambda_\psi$ are the dynamically generated axion
potential scales, and $f_\phi$ and $f_\psi$ are the axion decay constants. We expect
both $\Lambda_{\phi(\psi)}$ and $f_{\phi(\psi)}$ are smaller than $m_\mathrm{Pl}$,
especially the potential scale could be significantly so. There are many maxima,
minima and saddle points, and they lead to a very rich structure of the potential
depending on the values of the parameters. We illustrate a region of the potential
in Fig.~\ref{figpotential}

For simplicity, we concentrate on $f_\phi = f_\psi$ and $\Lambda_\phi =
\Lambda_\psi$, and occasionally we will omit the subscripts. Then from above we can
write the Hubble parameter as
\begin{equation}
H \simeq \sqrt{\frac{V}{3m_\mathrm{Pl}^2}} \sim \frac{\Lambda^2}{m_\mathrm{Pl}} \, .
\end{equation}
Near an extremum point, the effective mass is given by $|m| = 2\pi\Lambda^2/f$.
Hence
\begin{equation}\label{SReta}
\left| \frac{m}{H} \right| \sim \frac{m_\mathrm{Pl}}{f} \gtrsim 1 \, ,
\end{equation}
i.e. the mass is very heavy and therefore the slow-roll condition $\left|
m_\mathrm{Pl}^2V''/V \right| \ll 1$ is badly broken.

\section{Inflation}
\label{secinflation}

To compute the evolution of the scalar field on the potential given by
Eq.~(\ref{potential}), it is convenient to modify the equations of motion for $\phi$
and $\psi$ and to eliminate the scale factor $a(t)$. Using $d/dt = Hd/dN$, we can
derive the equations in terms of the number of $e$-folds $N$ as
\begin{equation}\label{fieldeq}
\phi_i'' + 3\phi_i' - \frac{\sum_j {\phi_j'}^2}{2m_\mathrm{Pl}^2}\phi_i' +
\frac{V_{,i}}{2V} \left( 6m_\mathrm{Pl}^2 - \sum_j {\phi_j'}^2 \right) = 0 \, ,
\end{equation}
where a prime denotes a derivative with respect to $N$, $V_{,i} = \partial
V/\partial\phi_i$ and the indices $i$ and $j$ denote the canonical
fields\footnote{For non-canonical fields, i.e.
\begin{equation*}
H^2 = \frac{1}{3m_\mathrm{Pl}^2} \left( \frac{1}{2}G_{ij}\dot\phi^i\dot\phi^j - V
\right) \, ,
\end{equation*}
where $G_{ij}$ is the metric of the field space, the equations are given by
\begin{equation*}
{\phi^i}'' + 3{\phi^i}' - \frac{G_{jk}{\phi^j}'{\phi^k}'}{2m_\mathrm{Pl}^2}{\phi^i}'
+ \Gamma^i_{jk}{\phi^j}'{\phi^k}' + \frac{G^{ij}V_{,j}}{2V} \left( 6m_\mathrm{Pl}^2
- G_{jk}{\phi^j}'{\phi^k}' \right) \, ,
\end{equation*}
where $\Gamma^i_{jk}$ is the Christoffel symbol constructed by $G_{ij}$.} $\phi$ and
$\psi$.

Now we study the trajectory of the field starting from a maximum of the
potential\footnote{We don't consider the inflationary phase which occurs as the
field rolls down from a saddle point. This \textit{saddle inflation} is extensively
studied in \cite{eks}.}. When the field is about to roll down the potential from a
maximum, we need to specify two quantities, namely, the displacement (how far is the
field off from the maximum?) and the misalignment (where is the field heading for?).
It is plausible to expect that due to quantum fluctuations the field is off the
maximum by an amplitude of $H \sim \Lambda^2/m_\mathrm{Pl}$. For the misalignment
$\theta$, which should be written as
\begin{equation}\label{ang}
\tan\theta = \frac{\psi}{\phi} \, ,
\end{equation}
first we note that near a maximum we have an approximate $U(1)$ symmetry. Hence we
can infer that as long as there is enough time for the field to get arbitrary
orientation, every direction is equivalent near a maximum. To estimate this time, we
note that the field is placed away from the maximum by $H$. Therefore, we can change
the question to ask how long it takes for the field to get randomly placed on a
circle of radius $H$. To have inflation, we require that
\begin{equation}
\frac{1}{2}\dot\phi^2 \lesssim V \lesssim 3m_\mathrm{Pl}^2H^2 \, ,
\end{equation}
and hence
\begin{equation}
\frac{1}{\dot\phi} \gtrsim \frac{1}{\sqrt{6}m_\mathrm{Pl}H} \, .
\end{equation}
Thus, we can estimate
\begin{equation}
\Delta N \gtrsim \frac{2\pi H^2}{\dot\phi} \sim \frac{H}{m_\mathrm{Pl}} \, ,
\end{equation}
i.e. the field is randomly placed almost instantaneously.

\begin{figure}[h]
\psfrag{N}{$N$}%
\psfrag{field}{field}%
\psfrag{phiN}{\scriptsize $\phi(N)$}%
\psfrag{psiN}{\scriptsize $\psi(N)$}%
\psfrag{phi}{$\phi$}%
\psfrag{psi}{$\psi$}%
\begin{center}
\includegraphics[width = 0.45\linewidth]{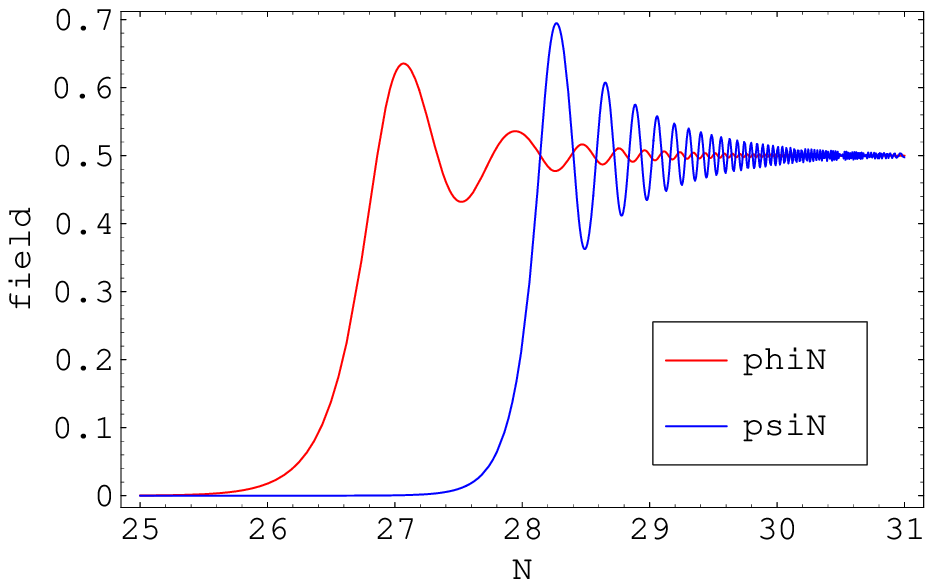}
\includegraphics[width = 0.45\linewidth]{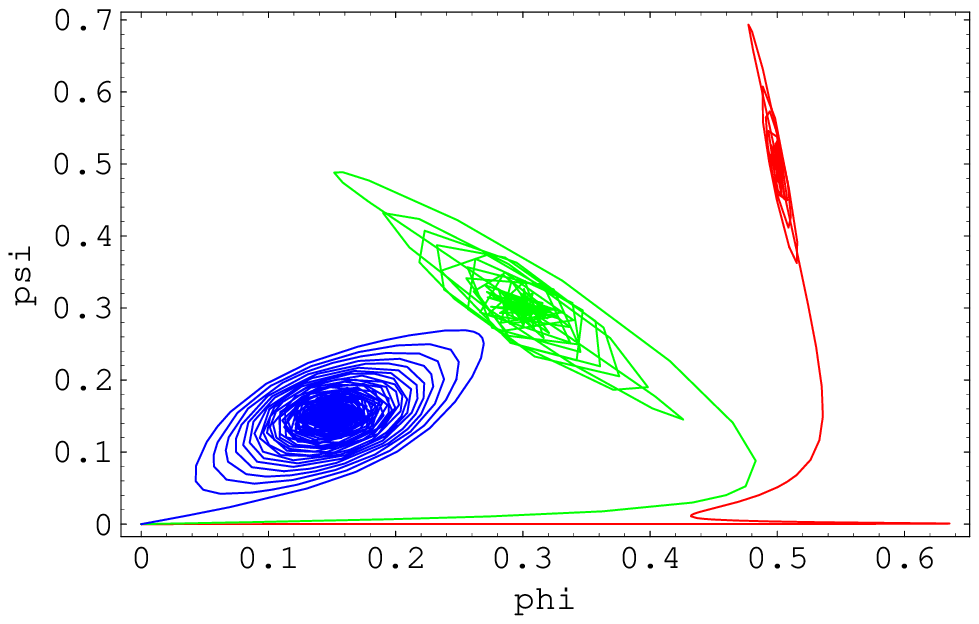}
\end{center}
\caption{(Left) evolution along $\phi$ and $\psi$ directions when $\Lambda =
10^{-10}m_\mathrm{Pl}$, $f = m_\mathrm{Pl}$, and $\theta = \pi/10000$, and (right)
several field trajectories for which the parameters are given by the same as the
left panel (red), $\Lambda = 10^{-9} m_\mathrm{Pl}$, $f = 2 \times 10^{-1}
m_\mathrm{Pl}$, and $\theta = \pi/100$ (green, magnified by 3 times), and $\Lambda =
10^{-8} m_\mathrm{Pl}$, $f = 5 \times 10^{-2} m_\mathrm{Pl}$, and $\theta = \pi/10$
(blue, magnified by 6 times).}%
\label{fieldfig}
\end{figure}

\begin{figure}[h]
\psfrag{logN}{$\log N$}%
\psfrag{logfmPl}{$\log(f/m_\mathrm{Pl})$}%
\psfrag{minus4}{\scriptsize $\Lambda = 10^{-4}m_\mathrm{Pl}$}%
\psfrag{minus6}{\scriptsize $\Lambda = 10^{-6}m_\mathrm{Pl}$}%
\psfrag{minus8}{\scriptsize $\Lambda = 10^{-8}m_\mathrm{Pl}$}%
\psfrag{minus10}{\scriptsize $\Lambda = 10^{-10}m_\mathrm{Pl}$}%
\begin{center}
\epsfig{file = 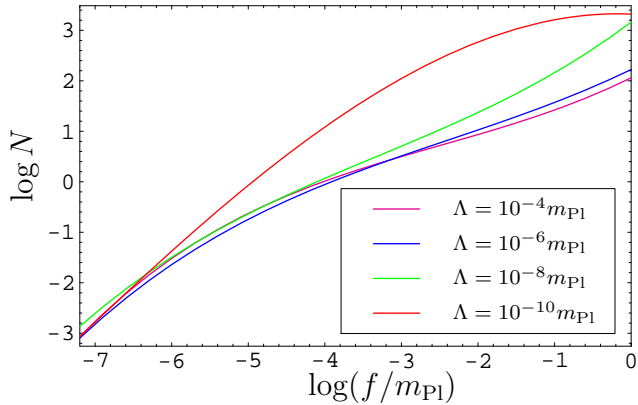, width = 9cm}%
\caption{Number of $e$-folds depending on $\Lambda$ and $f$, with $\theta =
\pi/100$.}%
\label{efoldfig}
\end{center}
\end{figure}

Some results of the numerical evolution of the field according to
Eq.~(\ref{fieldeq}) are shown in Fig.~\ref{fieldfig}. The left panel shows the field
evolution along $\phi$ and $\psi$ directions when $\Lambda = 10^{-10}
m_\mathrm{Pl}$, $f = m_\mathrm{Pl}$, and the initial misalignment $\theta =
\pi/10000$. Note that the initial misalignment is very small from $\psi = 0$, the
field rolls towards a saddle point and hence $\phi$ settles down at a minimum first.
This is clearly seen from several trajectories shown in the right panel of
Fig.~\ref{fieldfig}. Fig.~\ref{efoldfig} illustrates the dependence of the number of
$e$-folds on the energy scale $\Lambda$ and the decay constant $f$.

One interesting thing is that locked inflationary phase \cite{locked} hardly occurs.
This could be seen from the fact that there is no interaction term in the potential,
e.g., such as $\lambda\phi^2\psi^2$. Thus the oscillation of, say, $\phi$ is not
transmitted to the effective mass of $\psi$ to hold the field at a saddle point.
Note that even if we include an interaction term of the form
\begin{equation}
\frac{\Lambda_\phi^4\Lambda_\psi^4}{m_\mathrm{Pl}^4} \cos\left(
\frac{2\pi\phi}{f_\phi} \right) \cos\left( \frac{2\pi\psi}{f_\psi} \right) \, ,
\end{equation}
this term is suppressed by $m_\mathrm{Pl}^4$ and does not lead to any significant
change. Also we note that the dependence of the number of $e$-folds on the initial
misalignment $\theta$ \cite{ks} is very weak. We are considering $f \lesssim
m_\mathrm{Pl}$, and the slow-roll parameter $|\eta| \gtrsim \mathcal{O}(1)$, see
Eq.~(\ref{SReta}). Hence the field is very quickly rolling, and this makes the total
number of $e$-folds more or less the same irrespective of the initial misalignment
$\theta$. In Table~\ref{tabangdep}, we present several results depending on
$\theta$.

\begin{table}[h]
\begin{center}
\begin{tabular}{r||llllll}
\hline
    $\theta$ & $\pi/100$ & $\pi/40$ & $\pi/20$ & $\pi/10$ & $\pi/8$ & $\pi/5$
    \\
    $N$ & 30.4465 & 30.2241 & 30.2100 & 30.1144 & 30.3783 & 30.0687
    \\
\hline
\end{tabular}
\caption{Several results depending on different initial misalignment $\theta$.
Parameters are given by $\Lambda = 10^{-10} m_\mathrm{Pl}$ and $f = m_\mathrm{Pl}$.
Note that the number of $e$-folds varies less than 1.}
\end{center}
\label{tabangdep}
\end{table}

Now let us turn our attention to the perturbations produced during inflation. At the
early stage of inflation when the field is very close to a maximum, we can split the
field into the radial and angular components\footnote{In general, the field follows
curved and possibly complicated trajectory on the potential. If the number of
$e$-folds is large enough by e.g. some mechanism discussed in
Section~\ref{secremedies} and the perturbations produced when the field is around a
maximum is far outside the horizon, then this decomposition may complicate the
matter. In this case, one can just write \cite{deltaN}
\begin{equation*}
\mathcal{P} = \left( \frac{\partial N}{\partial\phi} \right)^2 \mathcal{P}_\phi +
\left( \frac{\partial N}{\partial\psi} \right)^2 \mathcal{P}_\psi
\end{equation*}
to calculate the power spectrum of the adiabatic perturbations after inflation.}
$\varphi$ and $\theta$, i.e.
\begin{equation}
\varphi = \sqrt{\phi^2 + \psi^2} \, ,
\end{equation}
and $\theta$ as Eq.~(\ref{ang}). Note that near a maximum, $U(1)$ symmetry is good
enough. The radial component $\varphi$ thus plays the role of the ``inflaton",
yielding adiabatic perturbations, while the angular component $\theta$ is orthogonal
to the field trajectory and its fluctuations correspond to the isocurvature
perturbations. Therefore the radial fluctuations follow the form of those from the
usual inverted quadratic potential \cite{sg2001}, but the angular ones are very
close to those of massless scalar field and the spectrum is nearly scale invariant
\cite{ks,bw}. Now, we can follow the $\delta N$ formalism \cite{deltaN} to write the
power spectrum of the primordial density perturbations as
\begin{align}
\mathcal{P} & = \sum_i \left( \frac{\partial N}{\partial\phi_i} \right)^2
\mathcal{P}_{\phi_i}
\nonumber \\
& = \left( \frac{\partial N}{\partial\varphi} \right)^2 \mathcal{P}_\varphi + \left(
\frac{\partial N}{\partial\theta} \right)^2 \mathcal{P}_\theta \, ,
\end{align}
where it is clear that both the adiabatic and the isocurvature components contribute
to the final perturbations. For the isocurvature perturbations to dominate the
primordial perturbations, we require that
\begin{equation}
\left( \frac{\partial N}{\partial\theta} \right)^2 \gg \left( \frac{\partial
N}{\partial\varphi} \right)^2 \, .
\end{equation}
Since the adiabatic component is boosted by a factor of $1/\sqrt{\epsilon}$,
generally we need $\left(\partial N/\partial\theta\right)^2 \gg 1$, i.e., the number
of $e$-folds is highly dependent on where the field is rolling towards. But as we
have seen in Table~\ref{tabangdep}, the number of $e$-folds is more or less the same
irrespective of the misalignment. Thus $(\partial N/\partial\theta)^2 \ll 1$ so the
contribution of the isocurvature component is negligible, and the density
perturbation spectrum is completely dominated by the contribution of the inflaton
component which is highly scale dependent.

Now let us consider the case when $f_\phi \neq f_\psi$, i.e. the potential is not
symmetric but ``squeezed''. Then, the field tends to roll along the squeezed
direction first irrespective of the initial misalignment, because that direction is
(much) steeper than the other one. Hence the isocurvature perturbations, i.e. the
perturbations orthogonal to the field trajectory, become irrelevant and the
``inflaton" perturbations are dominant. Let us assume that $\psi$ direction is
squeezed so that $f_\phi = \alpha f_\psi$ with $\alpha > 1$. With the rotated
slow-roll parameters
\begin{align}
\eta_{\sigma\sigma} & = \eta_{\phi\phi} \cos^2\theta + \eta_{\psi\psi} \sin^2\theta
\,
\nonumber \\
\eta_{ss} & = \eta_{\phi\phi} \sin^2\theta + \eta_{\psi\psi} \cos^2\theta \, ,
\end{align}
where $\sigma$ and $s$ denote the inflaton and the isocurvature components
respectively, the evolution of the isocurvature perturbations is given
by\footnote{Note that although this equation is valid under slow-roll approximation,
nevertheless it still gives physically interesting insights.} \cite{wbmr}
\begin{align}
\dot{\mathcal{S}} & \simeq (-2\epsilon + \eta_{\sigma\sigma} -
\eta_{ss})H\mathcal{S}
\nonumber \\
& = \left[ -2\epsilon + \frac{\alpha^2 - 1}{\alpha^2}
\frac{m_\mathrm{Pl}^2}{f_\psi^2} \left(\cos^2\theta - \sin^2\theta\right) \right]
H\mathcal{S}
\nonumber \\
& \simeq \left( -2\epsilon - \frac{\alpha^2 - 1}{\alpha^2}
\frac{m_\mathrm{Pl}^2}{f_\psi^2} \right) H\mathcal{S} \, ,
\end{align}
which is always negative. Intuitively, as a direction becomes more and more
squeezed, the potential becomes closer to the single field case where only the
inflaton perturbations exist.

\section{Remedies}
\label{secremedies}

In the previous section, we have seen that the number of $e$-folds obtained during
inflation driven by two decoupled string axion fields is generally not enough to
solve many cosmological problems. Also the power spectrum of density perturbations
is completely dominated by the radial contribution which is not scale invariant. The
naive inflation model of the two string axions is thus not sufficient and we need
some amendments to improve the situation. In this section we briefly summarise some
known remedies.

One obvious modification is to add more contribution to $H$ so that the field
receives greater friction and can roll down the potential slowly. One would be
tempted to add a positive cosmological constant, but this may cause two
difficulties. When the cosmological constant is too large, it may render the
universe to expand forever even after the field has settled down at a minimum. Also,
since the observed dark energy density is of $\mathcal{O} \left( 10^{-12}
\mathrm{eV}^4 \right)$, the magnitude of the cosmological constant one would like to
add is severely constrained. It is possible, however, to increase $H$ by adding more
fields and make them rolling slowly for many $e$-folds \cite{assisted}. Then
Eq.~(\ref{fieldeq}) becomes
\begin{equation}
\phi_i'' + 3\phi_i' - \frac{\sum_j{\phi_j'}^2}{2m_\mathrm{Pl}^2}\phi_i' +
\frac{V_{,i} \left( 6m_\mathrm{Pl}^2 - \sum_j{\phi_j'}^2 \right)}{2 \left( V +
3m_\mathrm{Pl}^2H_0^2 \right)} = 0 \, ,
\end{equation}
where the contribution $H_0$ by other fields is dependent on models: for example, if
we adopt N-flation\footnote{In a subsequent study \cite{endN}, we found that
although the detail depends on the underlying mass distribution, roughly 1/10 of the
total fields contribute at the end of the inflationary phase.} \cite{dkmw} and add
$k$ additional axions, we have
\begin{align}
3m_\mathrm{Pl}^2 H_0^2 & = V
\nonumber \\
& \simeq \sum_k \frac{1}{2}m_k^2\phi_k^2 \, ,
\end{align}
where we assume that each axion field is displaced not too far from the minimum. It
is possible that by some of the added fields further perturbations are generated
after inflation \cite{curvaton}, but the parameter space is heavily restricted for
this mechanism to work properly \cite{curvE}.

\begin{figure}[h]
\psfrag{logN}{$\log N$}%
\psfrag{logfmPl}{$\log(f/m_\mathrm{Pl})$}%
\psfrag{Pi4}{\scriptsize $\theta = \pi/4$}%
\psfrag{Pi6}{\scriptsize $\theta = \pi/6$}%
\psfrag{Pi10}{\scriptsize $\theta = \pi/10$}%
\psfrag{Pi100}{\scriptsize $\theta = \pi/100$}%
\begin{center}
\epsfig{file = 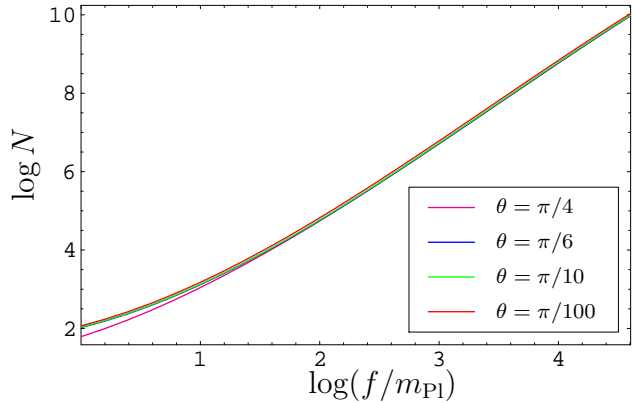, width = 9cm}%
\caption{Plot of $N$ versus $f$ and $\theta$. The energy scale $\Lambda$ is set to
be $10^{-4}m_\mathrm{Pl}$. We can see that as $f$ becomes bigger, we obtain larger
$N$ as expected. Also note that the dependence on $\theta$ is not significant.}%
\label{largeffig}
\end{center}
\end{figure}

Another clear way we can take is to make the axion decay constant $f$ larger than
$m_\mathrm{Pl}$ and follow the original wisdom of natural inflation \cite{natural}.
One difficulty we meet is that because it is required that $f \gg m_\mathrm{Pl}$ to
satisfy the slow-roll conditions, $f$ is outside the region of the effective field
theory description and is out of our control. An interesting alternative to build a
model where the above problem could be evaded is to identify the extra component
$A_5$ of a gauge field in a 5D theory with the inflaton \cite{extranatural}. Because
this model is based on higher dimensional theory, the effective decay constant
$f_\mathrm{eff} = (2\pi g_{4D} R)^{-1} \gg m_\mathrm{Pl}$, where $R$ is the radius
of the circle where the extra dimension is compactified, is reliable. Another way of
constructing $f > m_\mathrm{Pl}$ is to incorporate two different gauge groups,
making the effective decay constant larger than $m_\mathrm{Pl}$ by imposing a
symmetric relation between the couplings of the axions to the gauge groups
\cite{knp}. In Fig.~\ref{largeffig} several results of $e$-folds depending on $f$
and $\theta$ are shown. Note that when the field rolls towards a saddle point,
generally it acquires extra $e$-folds but the scale invariance of the perturbation
spectrum is broken when the field leaves the saddle point \cite{bsi}. To avoid this,
we should put strong constraints on the parameters \cite{eks}.

Also note that when $f \gg m_\mathrm{Pl}$, the minima are separated by a distance
far larger than $m_\mathrm{Pl}$. Hence, the extrema which intervene between those
minima become the sources of eternal topological inflation \cite{topological}. For
example, a saddle point which separates two minima along (say) $\psi$ direction, it
becomes a domain wall and in its vicinity we have an eternally inflating regime.

One fundamental question we can ask at this point is whether it is possible to find
regimes of moduli space where $f \gg m_\mathrm{Pl}$ in string theory. It seems that,
unfortunately, there exists no consistent scenario yet \cite{bdfg} although not all
the possibilities are explored. Building a consistent scenario where a large decay
constant could arise is still an interesting and challenging topic.

\section{Summary}
\label{summary}

We have investigated a simple inflation model which consists of two non-interacting
axions. The number of $e$-folds $N$ achieved while the field rolls off from a
maximum is dependent on both the energy scale $\Lambda$ and the decay constant $f$:
we obtain larger $N$ with smaller $\Lambda$ and with bigger $f$, but generally $N$
is not sufficient to solve cosmological problems. Interestingly, in spite of the
rich geometry of the potential, the initial misalignment $\theta$ hardly affects $N$
because of the absence of the interaction term, which also leads to no locked
inflationary phase. The perturbation spectrum is highly scale dependent, because the
scale invariant spectrum which arises from the isocurvature component is greatly
suppressed. To alleviate this situation, we may either increase $H$ by adding more
axion fields, or make $f$ bigger than $m_\mathrm{Pl}$ which is still unclear in the
context of string theory.

\subsection*{Acknowledgements}
I am grateful to James Cline and Richard Easther for helpful comments. It is a great
pleasure to thank Alexei Starobinsky for invaluable correspondences on the adiabatic
and isocurvature perturbations. I also thank the organisers of the 86th Les Houches
Summer Session where some part of this work was carried out.

\end{document}